\documentclass[aps,prl,reprint,longbibliography,floatfix]{revtex4-2}

\usepackage[utf8]{inputenc} 
\usepackage[T1]{fontenc} 
\usepackage{newtxtext,newtxmath} 
\usepackage[
  colorlinks=true,
  urlcolor=purple,
  citecolor=purple,
  filecolor=purple,
  linkcolor=purple
]{hyperref} 
\usepackage{amsmath, graphicx, xcolor} 

\begin{document}

\title{Complex complex landscapes}

\author{Jaron Kent-Dobias}
\author{Jorge Kurchan}

\affiliation{Laboratoire de Physique de l'Ecole Normale Supérieure, Paris, France}

\date\today

\begin{abstract}
  We study the saddle-points of the $p$-spin model -- the best understood
  example of a `complex' (rugged) landscape -- when its $N$ variables are
  complex. These points are the solutions to a system of $N$ random equations
  of degree $p-1$.  We solve for $\overline{\mathcal{N}}$, the number of
  solutions averaged over randomness in the $N\to\infty$ limit.  We find that
  it saturates the Bézout bound $\log\overline{\mathcal{N}}\sim N \log(p-1)$.
  The Hessian of each saddle is given by a random matrix of the form $C^\dagger
  C$, where $C$ is a complex symmetric Gaussian matrix with a shift to its diagonal.  Its
  spectrum has a transition where a gap develops that generalizes the notion of
  `threshold level' well-known in the real problem.  The results from the real
  problem are recovered in the limit of real parameters. In this case, only the
  square-root of the total number of solutions are real.  In terms of the
  complex energy, the solutions are divided into sectors where the saddles have
  different topological properties.
\end{abstract}

\maketitle

Spin-glasses have long been considered the paradigm of many variable `complex
landscapes,' a subject that includes neural networks and optimization problems,
most notably constraint satisfaction.  The most tractable family of these
are the mean-field spherical $p$-spin models \cite{Crisanti_1992_The} (for a
review see \cite{Castellani_2005_Spin-glass}) defined by the energy
\begin{equation} \label{eq:bare.hamiltonian}
  H_0 = \frac1{p!}\sum_{i_1\cdots i_p}^NJ_{i_1\cdots i_p}z_{i_1}\cdots z_{i_p},
\end{equation}
where $J$ is a symmetric tensor whose elements are real Gaussian variables and
$z\in\mathbb R^N$ is constrained to the sphere $z^2=N$. This problem has been
studied in the algebra \cite{Cartwright_2013_The} and probability literature
\cite{Auffinger_2012_Random, Auffinger_2013_Complexity}.  It has been attacked
from several angles: the replica trick to compute the Boltzmann--Gibbs
distribution \cite{Crisanti_1992_The}, a Kac--Rice \cite{Kac_1943_On,
Rice_1939_The, Fyodorov_2004_Complexity} procedure (similar to the
Fadeev--Popov integral) to compute the number of saddle-points of the energy
function \cite{Crisanti_1995_Thouless-Anderson-Palmer}, and the
gradient-descent---or more generally Langevin---dynamics staring from a
high-energy configuration \cite{Cugliandolo_1993_Analytical}. Thanks to the
simplicity of the energy, all these approaches yield analytic results in the
large-$N$ limit.

In this paper we extend the study to complex variables: we shall take
$z\in\mathbb C^N$ and $J$ to be a symmetric tensor whose elements are
\emph{complex} normal, with $\overline{|J|^2}=p!/2N^{p-1}$ and
$\overline{J^2}=\kappa\overline{|J|^2}$ for complex parameter $|\kappa|<1$. The
constraint remains $z^2=N$.

The motivations for this paper are of two types. On the practical side, there
are indeed situations in which complex variables appear naturally in disordered
problems: such is the case in which they are \emph{phases}, as in random laser
problems \cite{Antenucci_2015_Complex}. Quiver Hamiltonians---used to model
black hole horizons in the zero-temperature limit---also have a Hamiltonian
very close to ours \cite{Anninos_2016_Disordered}.  

There is however a more fundamental reason for this study: we know from
experience that extending a real problem to the complex plane often uncovers
underlying simplicity that is otherwise hidden. Consider, for example, the
procedure of starting from a simple, known Hamiltonian $H_{00}$ and studying
$\lambda H_{00} + (1-\lambda H_{0} )$, evolving adiabatically from $\lambda=0$
to $\lambda=1$, as is familiar from quantum annealing. The $H_{00}$ is a
polynomial of degree $p$  chosen to have simple, known saddles. Because we are
working in complex variables, and the saddles are simple all the way (we shall
confirm this), we may follow a single one from $\lambda=0$ to $\lambda=1$,
while  with real variables minima of functions appear and disappear, and this
procedure is not possible. The same idea may be implemented by performing
diffusion in the $J$s and following the roots, in complete analogy with Dyson's
stochastic dynamics \cite{Dyson_1962_A}.

The spherical constraint is enforced using the method of Lagrange multipliers:
introducing $\epsilon\in\mathbb C$, our energy is
\begin{equation} \label{eq:constrained.hamiltonian}
  H = H_0+\frac p2\epsilon\left(N-\sum_i^Nz_i^2\right).
\end{equation}
 We choose to
constrain our model by $z^2=N$ rather than $|z|^2=N$ in order to preserve the
analyticity of $H$. The nonholomorphic constraint also has a disturbing lack of
critical points nearly everywhere: if $H$ were so constrained, then
$0=\partial^* H=-p\epsilon z$ would only be satisfied for $\epsilon=0$.  

The critical points are of $H$ given by the solutions to the set of equations
\begin{equation} \label{eq:polynomial}
  \frac{p}{p!}\sum_{j_1\cdots j_{p-1}}^NJ_{ij_1\cdots j_{p-1}}z_{j_1}\cdots z_{j_{p-1}}
  = p\epsilon z_i
\end{equation}
for all $i=\{1,\ldots,N\}$, which for fixed $\epsilon$ is a set of $N$
equations of degree $p-1$, to which one must add the constraint. 
In this sense
this study also provides a complement to the work on the distribution of zeroes
of random polynomials \cite{Bogomolny_1992_Distribution}, which are for $N=1$
and $p\to\infty$.
We see from \eqref{eq:polynomial} that at any critical point, $\epsilon=H/N$, the average energy.

Since $H$ is holomorphic, any critical point of $\operatorname{Re}H$ is also a
critical point of $\operatorname{Im}H$. The number of critical points of $H$ is
therefore the same as that of $\operatorname{Re}H$. From each saddle 
emerge gradient lines of $\operatorname{Re}H$, which are also ones of constant
$\operatorname{Im}H$ and therefore constant phase.

Writing $z=x+iy$, $\operatorname{Re}H$ can be considered a real-valued function
of $2N$ real variables. Its number of saddle-points is given by the usual
Kac--Rice formula:
\begin{equation} \label{eq:real.kac-rice}
  \begin{aligned}
    \mathcal N_J&(\kappa,\epsilon)
      = \int dx\,dy\,\delta(\partial_x\operatorname{Re}H)\delta(\partial_y\operatorname{Re}H) \\
      &\hspace{6pc}\times\left|\det\begin{bmatrix}
          \partial_x\partial_x\operatorname{Re}H & \partial_x\partial_y\operatorname{Re}H \\
          \partial_y\partial_x\operatorname{Re}H & \partial_y\partial_y\operatorname{Re}H
        \end{bmatrix}\right|.
  \end{aligned}
\end{equation}
The Cauchy--Riemann equations may be used to write this in a manifestly complex
way.  With the Wirtinger derivative $\partial=\frac12(\partial_x-i\partial_y)$,
one can write $\partial_x\operatorname{Re}H=\operatorname{Re}\partial H$ and
$\partial_y\operatorname{Re}H=-\operatorname{Im}\partial H$. Carrying these
transformations through, we have
\begin{equation} \label{eq:complex.kac-rice}
  \begin{aligned}
    \mathcal N_J&(\kappa,\epsilon)
      = \int dx\,dy\,\delta(\operatorname{Re}\partial H)\delta(\operatorname{Im}\partial H) \\
                &\hspace{6pc}\times\left|\det\begin{bmatrix}
            \operatorname{Re}\partial\partial H & -\operatorname{Im}\partial\partial H \\
            -\operatorname{Im}\partial\partial H & -\operatorname{Re}\partial\partial H
          \end{bmatrix}\right| \\
      &= \int dx\,dy\,\delta(\operatorname{Re}\partial H)\delta(\operatorname{Im}\partial H)
        \left|\det[(\partial\partial H)^\dagger\partial\partial H]\right| \\
      &= \int dx\,dy\,\delta(\operatorname{Re}\partial H)\delta(\operatorname{Im}\partial H)
        |\det\partial\partial H|^2.
  \end{aligned}
\end{equation}
This gives three equivalent expressions for the determinant of the Hessian: as
that of a $2N\times 2N$ real matrix, that of an $N\times N$ Hermitian matrix,
i.e. the norm squared of that of an $N\times N$ complex symmetric matrix.

These equivalences belie a deeper connection between the spectra of the
corresponding matrices. Each positive eigenvalue of the real matrix has a
negative partner. For each such pair $\pm\lambda$, $\lambda^2$ is an eigenvalue
of the Hermitian matrix and $|\lambda|$ is a \emph{singular value} of the
complex symmetric matrix. The distribution of positive eigenvalues of the
Hessian is therefore the same as the distribution of singular values of
$\partial\partial H$, or the distribution of square-rooted eigenvalues of
$(\partial\partial H)^\dagger\partial\partial H$.

The expression \eqref{eq:complex.kac-rice} is to be averaged over $J$ to give
the complexity $\Sigma$ as $N \Sigma= \overline{\log\mathcal N} = \int dJ \,
\log \mathcal N_J$, a calculation that involves the replica trick. In most the
parameter-space that we shall study here, the \emph{annealed approximation} $N
\Sigma \sim \log \overline{ \mathcal N} = \log\int dJ \, \mathcal N_J$ is
exact. 

A useful property of the Gaussian $J$ is that gradient and Hessian at fixed
$\epsilon$ are statistically independent \cite{Bray_2007_Statistics,
Fyodorov_2004_Complexity}, so that the $\delta$-functions and the Hessian may
be averaged independently. The $\delta$-functions are converted to exponentials
by the introduction of auxiliary fields $\hat z=\hat x+i\hat y$.  The average
of those factors over $J$ can then be performed. A generalized
Hubbard--Stratonovich allows a change of variables from the $4N$ original
and auxiliary fields to eight bilinears defined by $Na=|z|^2$, $N\hat a=|\hat
z|^2$, $N\hat c=\hat z^2$, $Nb=\hat z^*z$, and $Nd=\hat zz$ (and their
conjugates). The result, to leading order in $N$, is
\begin{equation} \label{eq:saddle}
    \overline{\mathcal N}(\kappa,\epsilon)
        = \int da\,d\hat a\,db\,db^*d\hat c\,d\hat c^*dd\,dd^*e^{Nf(a,\hat a,b,\hat c,d)},
\end{equation}
where the argument of the exponential is
\begin{widetext}
  \begin{equation}
    f=2+\frac12\log\det\frac12\begin{bmatrix}
      1 & a & d & b \\
      a & 1 & b^* & d^* \\
      d & b^* & \hat c & \hat a \\
      b & d^* & \hat a & \hat c^*
    \end{bmatrix}
    +\int d\lambda\,\rho(\lambda)\log|\lambda|^2
    +p\operatorname{Re}\left\{
      \frac18\left[\hat aa^{p-1}+(p-1)|d|^2a^{p-2}+\kappa(\hat c^*+(p-1)b^2)\right]-\epsilon b
    \right\}.
  \end{equation}
  The integral of the distribution $\rho$ of eigenvalues of $\partial\partial
  H$ comes from the Hessian and is dependant on $a$ alone. This function has an
  extremum in $\hat a$, $b$, $\hat c$, and $d$ at which its value is
  \begin{equation} \label{eq:free.energy.a}
      f(a)=1+\frac12\log\left(\frac4{p^2}\frac{a^2-1}{a^{2(p-1)}-|\kappa|^2}\right)+\int d\lambda\,\rho(\lambda)\log|\lambda|^2
      -2C_+[\operatorname{Re}(\epsilon e^{-i\theta})]^2-2C_-[\operatorname{Im}(\epsilon e^{-i\theta})]^2,
  \end{equation}
\end{widetext}
where $\theta=\frac12\arg\kappa$ and
\begin{equation}
  C_{\pm}=\frac{a^p(1+p(a^2-1))\mp a^2|\kappa|}{a^{2p}\pm(p-1)a^p(a^2-1)|\kappa|-a^2|\kappa|^2}.
\end{equation}
This leaves a single parameter, $a$, which dictates the magnitude of $|z|^2$,
or alternatively the magnitude $y^2$ of the imaginary part. The latter vanishes
as $a\to1$, where (as we shall see) one recovers known results for the real
$p$-spin.

\begin{figure}[b]
  \centering

  \includegraphics{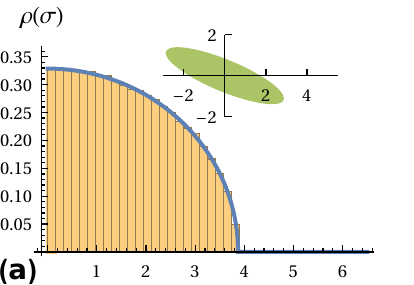}
  \includegraphics{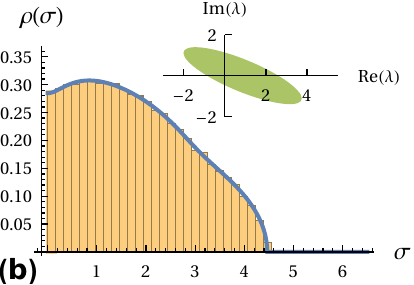}\\
  \includegraphics{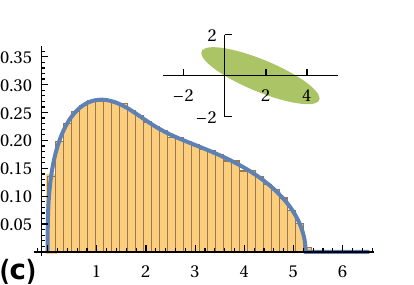}
  \includegraphics{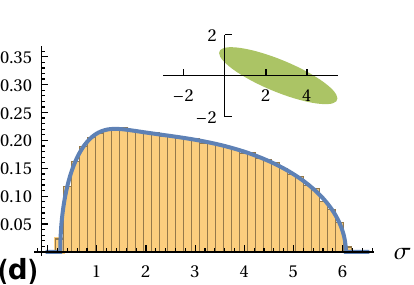}

  \caption{
    Eigenvalue and singular value spectra of the matrix $\partial\partial H$
    for $p=3$, $a=\frac54$, and $\kappa=\frac34e^{-i3\pi/4}$ with (a)
    $\epsilon=0$, (b) $\epsilon=-\frac12|\epsilon_{\mathrm{th}}|$, (c)
    $\epsilon=-|\epsilon_{\mathrm{th}}|$, and (d)
    $\epsilon=-\frac32|\epsilon_{\mathrm{th}}|$. The shaded region of each inset
    shows the support of the eigenvalue distribution. The solid line on each
    plot shows the distribution of singular values, while the overlaid
    histogram shows the empirical distribution from $2^{10}\times2^{10}$ complex
    normal matrices with the same covariance and diagonal shift as
    $\partial\partial H$.
  } \label{fig:spectra}
\end{figure}

The Hessian $\partial\partial H=\partial\partial H_0-p\epsilon I$ is equal to
the unconstrained Hessian with a constant added to its diagonal. The eigenvalue
distribution $\rho$ is therefore related to the unconstrained distribution
$\rho_0$ by a similar shift: $\rho(\lambda)=\rho_0(\lambda+p\epsilon)$. The
Hessian of the unconstrained Hamiltonian is
\begin{equation} \label{eq:bare.hessian}
  \partial_i\partial_jH_0
  =\frac{p(p-1)}{p!}\sum_{k_1\cdots k_{p-2}}^NJ_{ijk_1\cdots k_{p-2}}z_{k_1}\cdots z_{k_{p-2}},
\end{equation}
which makes its ensemble that of Gaussian complex symmetric matrices, when the
direction along the constraint is neglected. Given its variances
$\overline{|\partial_i\partial_j H_0|^2}=p(p-1)a^{p-2}/2N$ and
$\overline{(\partial_i\partial_j H_0)^2}=p(p-1)\kappa/2N$, $\rho_0(\lambda)$ is
constant inside the ellipse
\begin{equation} \label{eq:ellipse}
  \left(\frac{\operatorname{Re}(\lambda e^{i\theta})}{a^{p-2}+|\kappa|}\right)^2+
  \left(\frac{\operatorname{Im}(\lambda e^{i\theta})}{a^{p-2}-|\kappa|}\right)^2
  <\frac{p(p-1)}{2a^{p-2}}
\end{equation}
where $\theta=\frac12\arg\kappa$ \cite{Nguyen_2014_The}. The eigenvalue
spectrum of $\partial\partial H$ is therefore constant inside the same ellipse
translated so that its center lies at $-p\epsilon$.  Examples of these
distributions are shown in the insets of Fig.~\ref{fig:spectra}.

The eigenvalue spectrum of the Hessian of the real part is different from the
spectrum $\rho(\lambda)$ of $\partial\partial H$, but rather equivalent to the
square-root eigenvalue spectrum of $(\partial\partial H)^\dagger\partial\partial H$;
in other words, the singular value spectrum $\rho(\sigma)$ of $\partial\partial
H$. When $\kappa=0$ and the elements of $J$ are standard complex normal, this
is a complex Wishart distribution. For $\kappa\neq0$ the problem changes, and
to our knowledge a closed form is not in the literature.  We have worked out an
implicit form for this spectrum using the replica method.

Introducing replicas to bring the partition function into the numerator of the
Green function \cite{Livan_2018_Introduction} gives
\begin{widetext}
  \begin{equation} \label{eq:green.replicas}
    G(\sigma)=\lim_{n\to0}\int d\zeta\,d\zeta^*\,(\zeta_i^{(0)})^*\zeta_i^{(0)}
      \exp\left\{
      \frac12\sum_\alpha^n\left[(\zeta_i^{(\alpha)})^*\zeta_i^{(\alpha)}\sigma
        -\operatorname{Re}\left(\zeta_i^{(\alpha)}\zeta_j^{(\alpha)}\partial_i\partial_jH\right)
      \right]
    \right\},
  \end{equation}
  with sums taken over repeated Latin indices.  The average is then made over
  $J$ and Hubbard--Stratonovich is used to change variables to the replica matrices
  $N\alpha_{\alpha\beta}=(\zeta^{(\alpha)})^*\cdot\zeta^{(\beta)}$ and
  $N\chi_{\alpha\beta}=\zeta^{(\alpha)}\cdot\zeta^{(\beta)}$ and a series of
  replica vectors. The replica-symmetric ansatz leaves all off-diagonal
  elements and vectors zero, and
  $\alpha_{\alpha\beta}=\alpha_0\delta_{\alpha\beta}$,
  $\chi_{\alpha\beta}=\chi_0\delta_{\alpha\beta}$. The result is
  \begin{equation}\label{eq:green.saddle}
    \overline G(\sigma)=N\lim_{n\to0}\int d\alpha_0\,d\chi_0\,d\chi_0^*\,\alpha_0
    \exp\left\{nN\left[
      1+\frac{p(p-1)}{16}a^{p-2}\alpha_0^2-\frac{\alpha_0\sigma}2+\frac12\log(\alpha_0^2-|\chi_0|^2)
      +\frac p4\operatorname{Re}\left(\frac{(p-1)}8\kappa^*\chi_0^2-\epsilon^*\chi_0\right)
    \right]\right\}.
    \nonumber 
  \end{equation}
\end{widetext}

The argument of the exponential has several saddles. The solutions $\alpha_0$
are the roots of a sixth-order polynomial, and the root with the smallest value
of $\operatorname{Re}\alpha_0$ in all the cases we studied gives the correct
solution. A detailed analysis of the saddle point integration is needed to
understand why this is so. Given such $\alpha_0$, the density of singular
values follows from the jump across the cut, or
\begin{equation} \label{eq:spectral.density}
  \rho(\sigma)=\frac1{i\pi N}\left(
    \lim_{\operatorname{Im}\sigma\to0^+}\overline G(\sigma)
    -\lim_{\operatorname{Im}\sigma\to0^-}\overline G(\sigma)
  \right)
\end{equation}
Examples can be seen in Fig.~\ref{fig:spectra} compared with numeric
experiments.

The transition from a one-cut to two-cut singular value spectrum naturally
corresponds to the origin leaving the support of the eigenvalue spectrum.
Weyl's theorem requires that the product over the norm of all eigenvalues must
not be greater than the product over all singular values \cite{Weyl_1912_Das}.
Therefore, the absence of zero eigenvalues implies the absence of zero singular
values. The determination of the threshold energy -- the energy at which the
distribution of singular values becomes gapped -- is then reduced to a
geometry problem, and yields
\begin{equation} \label{eq:threshold.energy}
  |\epsilon_{\mathrm{th}}|^2
  =\frac{p-1}{2p}\frac{(1-|\delta|^2)^2a^{p-2}}
  {1+|\delta|^2-2|\delta|\cos(\arg\kappa+2\arg\epsilon)}
\end{equation}
for $\delta=\kappa a^{-(p-2)}$.

Given $\rho$, the integral in \eqref{eq:free.energy.a} may be preformed for
arbitrary $a$. The resulting expression is maximized for $a=\infty$ for all
values of $\kappa$ and $\epsilon$. Taking this saddle gives
\begin{equation} \label{eq:bezout}
  \log\overline{\mathcal N}(\kappa,\epsilon)
  =N\log(p-1).
\end{equation}
This is, to this order,  precisely the Bézout bound, the maximum number of
solutions that $N$ equations of degree $p-1$ may have
\cite{Bezout_1779_Theorie}. That we saturate this bound is perhaps not
surprising, since the coefficients of our polynomial equations
\eqref{eq:polynomial} are complex and have no symmetries. Reaching Bézout in
\eqref{eq:bezout} is not our main result, but it provides a good check.
Analogous asymptotic scaling has been found for the number of pure Higgs states
in supersymmetric quiver theories \cite{Manschot_2012_From}.

More insight is gained by looking at the count as a function of $a$, defined by
$\overline{\mathcal N}(\kappa,\epsilon,a)=e^{Nf(a)}$.  In the large-$N$ limit,
this is the cumulative number of critical points, or the number of critical
points $z$ for which $|z|^2\leq a$. We likewise define the $a$-dependant
complexity $\Sigma(\kappa,\epsilon,a)=N\log\overline{\mathcal
N}(\kappa,\epsilon,a)$

\begin{figure}[htpb]
  \centering
  \includegraphics{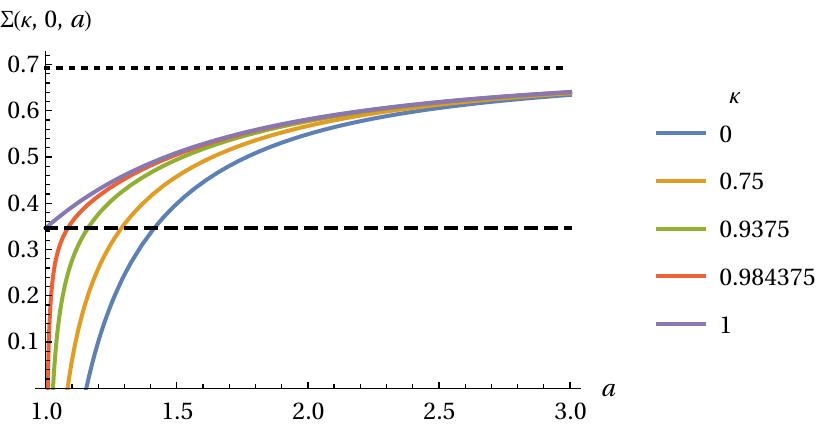}
  \caption{
    The complexity of the 3-spin model at $\epsilon=0$ as a function of
    $a=|z|^2=1+y^2$ at several values of $\kappa$. The dashed line shows
    $\frac12\log(p-1)$, while the dotted shows $\log(p-1)$.
  } \label{fig:complexity}
\end{figure}

Everything is analytically tractable for $\epsilon=0$, giving
\begin{equation} \label{eq:complexity.zero.energy}
  \Sigma(\kappa,0,a)
  =\log(p-1)-\frac12\log\left(\frac{1-|\kappa|^2a^{-2(p-1)}}{1-a^{-2}}\right).
\end{equation}
Notice that the limit of this expression as $a\to\infty$ corresponds with
\eqref{eq:bezout}, as expected. This is plotted as a function of $a$ for
several values of $\kappa$ in Fig.~\ref{fig:complexity}. For any $|\kappa|<1$,
the complexity goes to negative infinity as $a\to1$, i.e., as the spins are
restricted to the reals.  This is natural, given that the $y$ contribution to
the volume shrinks to zero as that of an $N$-dimensional sphere $\sum_i y_i^2=N(a-1)$ with volume
$\sim(a-1)^N$.  However, when the result is analytically continued to
$\kappa=1$ (which corresponds to real $J$) something novel occurs: the
complexity has a finite value at $a=1$. Since the $a$-dependence gives a
cumulative count, this implies a $\delta$-function density of critical points
along the line $y=0$.  The number of critical points contained within is
\begin{equation}
  \lim_{a\to1}\lim_{\kappa\to1}\log\overline{\mathcal N}(\kappa,0,a)
  = \frac12N\log(p-1),
\end{equation}
half of \eqref{eq:bezout} and corresponding precisely to the number of critical
points of the real $p$-spin model (note the role of conjugation symmetry,
already underlined in \cite{Bogomolny_1992_Distribution}). The full
$\epsilon$-dependence of the real $p$-spin is recovered by this limit as
$\epsilon$ is varied.

\begin{figure}[b]
  \centering
  \includegraphics{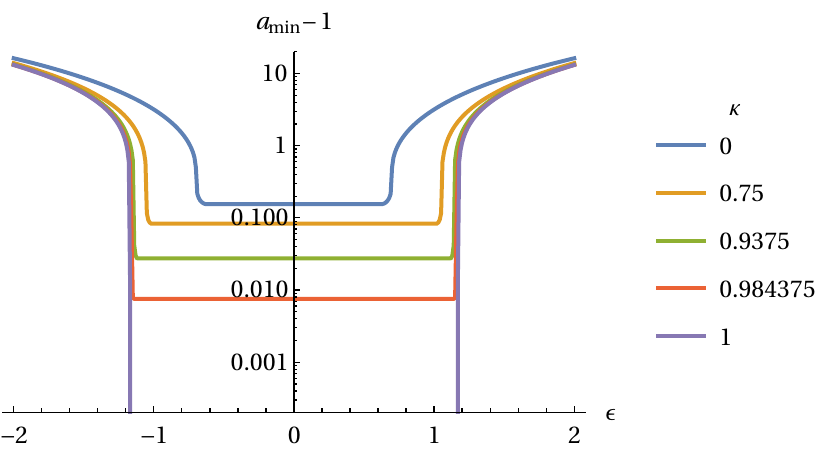}
  \caption{
    The minimum value of $a$ for which the complexity is positive as a function
    of (real) energy $\epsilon$  for the 3-spin model at several values of
    $\kappa$.
  } \label{fig:desert}
\end{figure}

These qualitative features carry over to nonzero $\epsilon$. In
Fig.~\ref{fig:desert} we show that for $\kappa<1$ there is always a gap of $a$
close to one for which there are no solutions. When $\kappa=1$---the analytic
continuation to the real computation---the situation is more interesting. In
the range of energies where there are real solutions this gap closes, which is
only possible if the density of solutions diverges at $a=1$.  Another
remarkable feature of this limit is that there is still a gap without solutions
around `deep' real energies where there is no real solution. A moment's thought
tells us that this is a necessity: otherwise a small perturbation of the $J$s
could produce an unusually deep solution to the real problem, in a region where
this should not happen.

\begin{figure}[t]
  \centering

  \includegraphics{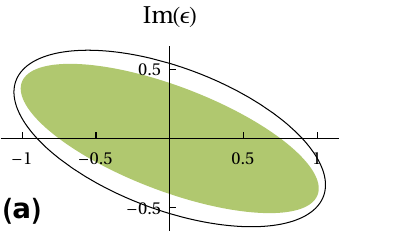}
  \includegraphics{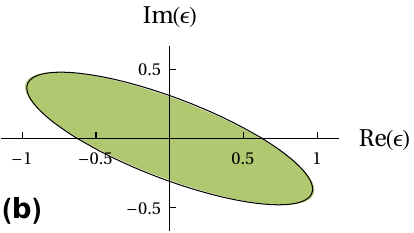} \\
  \includegraphics{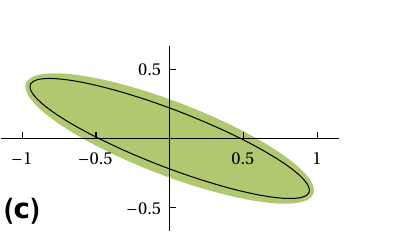}
  \includegraphics{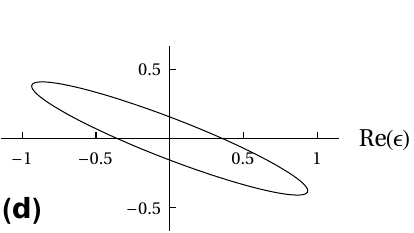}

  \caption{
    Energies at which states exist (green shaded region) and threshold energies
    (black solid line) for the 3-spin model with
    $\kappa=\frac34e^{-i3\pi/4}$ and (a) $a=2$, (b) $a=1.325$, (c) $a=1.125$,
    and (d) $a=1$. No shaded region is shown in (d) because no states exist at
    any energy.
  } \label{fig:eggs}
\end{figure}

The relationship between the threshold and ground, or extremal, state energies
is richer than in the real case. In Fig.~\ref{fig:eggs} these are shown in the
complex-$\epsilon$ plane for several examples. Depending on the parameters, the
threshold might always come at smaller magnitude than the extremal state, or
always come at larger magnitude, or cross as a function of complex argument.
For sufficiently large $a$ the threshold always comes at larger magnitude than
the extremal state. If this were to happen in the real case, it would likely
imply our replica symmetric computation is unstable, since having a ground
state above the threshold implies a ground state Hessian with many negative
eigenvalues, a contradiction. However, this is not an obvious contradiction in
the complex case. The relationship between the threshold, i.e., where the gap
appears, and the dynamics of, e.g., a minimization algorithm or physical
dynamics, are a problem we hope to address in future work.

This paper provides a first step for the study of a complex landscape with
complex variables. The next obvious one is to study the topology of the
critical points and gradient lines of constant phase.  We anticipate that the
threshold level, where the system develops a mid-spectrum gap, will play a
crucial role as it does in the real case.

\begin{acknowledgments}
  We wish to thank Alexander Altland, Satya Majumdar and Gregory Schehr for a useful suggestions.
  JK-D and JK are supported by the Simons Foundation Grant No.~454943.
\end{acknowledgments}

\bibliographystyle{apsrev4-2}
\bibliography{bezout}

\end{document}